\documentstyle[aps]{revtex}
\tighten
\begin{document}
\draft


\preprint{hep-th/0007239}
\title{Quantum effects, brane tension and large hierarchy in the
brane-world}
\author{Shinji Mukohyama}
\address{
Department of Physics and Astronomy, University of Victoria\\ 
Victoria, BC, Canada V8W 3P6
}
\date{\today}

\maketitle


\begin{abstract} 

Semiclassical Einstein's equation in five-dimension with a negative
cosmological constant and conformally invariant bulk matter fields is
examined in the brane world scenario with the $S^1/Z_2$
compactification. When numbers $N_{b,f}$ of bosonic and fermionic
fields satisfy $I<\kappa^2l^{-3}(N_b-N_f)<32I$, we obtain an exact
semiclassical solution which has two static branes with positive
tension and for which the warp factor can be arbitrarily large. Here,
$\kappa^2$ is the five-dimensional gravitational constant, $l$ is a
length scale determined by the negative five-dimensional cosmological
constant, and $I$ is a dimensionless positive constant of order
unity. However, in order to obtain a large warp factor, fine tuning of
brane tensions is required. Hence, in order to solve the hierarchy
problem, we need to justify the fine tuning by another new mechanism. 

\end{abstract}



Many unified theories require spacetime dimensionality higher than
four. For example, superstring theory and M-theory require ten and
eleven dimensions, respectively~\cite{Polchinski}. Since our observed
universe is four-dimensional, the spacetime dimensions should be 
reduced in such theories. One of the methods is called Kaluza-Klein 
compactification~\cite{KK}. An alternative method was recently
proposed by Randall and Sundrum~\cite{RS1,RS2}, and it consists of two
similar but distinct scenarios. In the first scenario~\cite{RS1}, they
considered a three-brane with negative tension as our universe and
showed that the hierarchy problem may be solved by a large redshift
factor, which is usually called a warp factor, between our brane and
another brane called a hidden brane. In the second
scenario~\cite{RS2}, a three-brane with positive tension is 
considered as our universe and four-dimensional Newton's law can be
realized on the brane. So far, many works were done on various aspects
of these brane-world scenarios: effective four-dimensional Einstein's 
equation on a positive tension brane~\cite{SMS}, weak
gravity~\cite{gravity,gravity-stabilization}, black
holes~\cite{Black-hole}, inflating branes~\cite{Inflating-brane},
cosmologies~\cite{Cosmology1,Cosmology2,Friedmann,Perturbation,creation},
and so on.

However, the solution of the hierarchy problem was not attempted in
the second scenario while four-dimensional Newton's law was derived in
the modified first scenario with radion
stabilization~\cite{gravity-stabilization}. Therefore, it might be 
concluded that we have to assume that our brane has negative tension
in order to obtain large hierarchy. The purpose of this paper is to
suggest that this may not be necessarily the case if a quantum effect
called Casimir effect is taken into account. (Refs.~\cite{GPT,Casimir}
also discussed Casimir effect in the brane world scenario.) To be
precise, we seek an exact semiclassical solution which has two static
branes with positive tension and for which the warp factor can be
arbitrarily large. (Refs.~\cite{2-positive-branes} also investigated
models with two positive tension branes.) In this paper such a
solution is found when numbers $N_{b,f}$ of bosonic and fermionic
fields satisfy  $I<\kappa^2l^{-3}(N_b-N_f)<32I$, where $\kappa^2$ is
the five-dimensional gravitational constant, $l$ is a length scale 
determined by the negative five-dimensional cosmological constant, and
$I$ is a dimensionless constant of order unity. However, in order to
obtain a large warp factor, fine tuning of brane tensions is
required.

This paper is organized as follows. First, we seek a static solution
of semiclassical Einstein's equation including Casimir effect due to
conformally invariant bulk fields. An explicit expression for the warp 
factor is given. Next, we extend the result to inflating
branes and discuss about the cosmological constant problem in this
model. Finally, we summarize this paper and discussions are given for
the hierarchy problem, extension to general homogeneous, isotropic
branes, and stability against perturbations.



To begin with, we consider a five-dimensional metric of the form 
%
\begin{equation}
 g_{MN}dx^Mdx^N = 
	e^{-2\alpha(w)}\eta_{\mu\nu}dx^{\mu}dx^{\nu} + dw^2, 
	\label{eqn:metric-ansatz}
\end{equation}
where $\eta_{\mu\nu}$ represents the four-dimensional
Minkowski metric. This is a general five-dimensional metric with
the four-dimensional Poincar\'{e} invariance. With this metric, we
analyze the semiclassical Einstein equation 
$G_{MN}+\Lambda g_{MN}=\kappa^2T_{MN}$ to take into account quantum
effects, where $T_{MN}$ should be understood as an expectation value
of energy momentum tensor. Hereafter, we assume that $\Lambda<0$ as in 
refs.~\cite{RS1,RS2} and, for simplicity, we consider quantized
conformally invariant fields as fields contributing to $T_{MN}$. In
this case, since there is no trace anomaly in odd dimension~\cite{BD},
we have $T^M_M=0$ and thus $R+20l^{-2}=0$, where $l$ is defined by
$\Lambda=-6l^{-2}$. This equation is written as
%
\begin{equation}
 (\alpha')^2 = \frac{2}{5}\alpha'' +l^{-2}, 
\end{equation}
and solved analytically to give
%
\begin{eqnarray}
 e^{-5\alpha} & = &C_1
	\cosh^2[\frac{5}{2}l^{-1}(w-w_0)],\quad \mbox{or}
	\label{eqn:bulk-sol-1}\\
 e^{-5\alpha} & = & C_2
	\sinh^2[\frac{5}{2}l^{-1}(w-w_0)],\quad \mbox{or}
	\label{eqn:bulk-sol-2}\\
 \alpha & = & \pm l^{-1}(w-w_0), 
	\label{eqn:bulk-sol-3}
\end{eqnarray}
where $C_{1,2}$ are positive constants. Correspondingly, energy
momentum tensor should be of the following form determined by
$T_{MN}=\kappa^{-2}(G_{MN}+\Lambda g_{MN})$. 
%
\begin{equation}
 T_{MN}dx^Mdx^N = \kappa^{-2}l^{-2}{\cal N}e^{5\alpha(w)}
	(e^{-2\alpha(w)}\eta_{\mu\nu}dx^{\mu}dx^{\nu} - 4dw^2),
	\label{eqn:bulk-energy-momentum}
\end{equation}
where ${\cal N}=\frac{3}{2}C_1(>0)$, $-\frac{3}{2}C_2(<0)$, and $0$
for the solutions (\ref{eqn:bulk-sol-1}), (\ref{eqn:bulk-sol-2}), and 
(\ref{eqn:bulk-sol-3}), respectively. It is easily seen from
(\ref{eqn:bulk-sol-3}) that the pure anti-deSitter spacetime used in
refs.~\cite{RS1,RS2} is recovered for ${\cal N}=0$.


Now let us compactify the $w$-direction by $S^1/Z_2$: we adopt the
identification $w\sim w+2L$, $w\sim -w$, where $L$ is a constant
representing the distance between two fixed points of the
$S^1/Z_2$. This corresponds to a configuration including a $3$-brane
at $w=0$ and another $3$-brane at $w=L$. By denoting tensions of these
branes by $\lambda$ and $\bar{\lambda}$, respectively, Israel's
junction conditions~\cite{Israel} at $w=0$ and $w=L$ are written as 
%
\begin{equation}
 \lambda = 6\kappa^{-2}\alpha'(0) , \quad
 \bar{\lambda} = -6\kappa^{-2}\alpha'(L). 
	\label{eqn:Israel}
\end{equation}
For non-zero ${\cal N}$, $\alpha'$ for above solutions is
independent of the absolute value of the constant ${\cal N}$, and
these junction conditions are sufficient to determine the constant
$w_0$ in (\ref{eqn:bulk-sol-1}) or (\ref{eqn:bulk-sol-2}) and 
the distance $L$. On the other hand, regarding an actual value of
${\cal N}$, because of the compactness of the $w$-direction and the
existence of the boundaries, it is expected that ${\cal N}\ne 0$ in
general. Actually, in ref.~\cite{GPT} Garriga et.al calculated energy
momentum tensor for both bosonic and fermionic fields in the
background (\ref{eqn:metric-ansatz}) and showed the form
(\ref{eqn:bulk-energy-momentum}) with ${\cal N}$ given by 
%
\begin{equation}
 \kappa^{-2}l^{-2}{\cal N} = \frac{1}{2}A(N_b-N_f)
	\left[\int_0^Le^{\alpha(w)}dw\right]^{-5}, 
	\label{eqn:constraint}
\end{equation}
where $A$ is a positive constant given by $A=\pi^2\zeta'_R(-4)/32$ and 
$N_{b,f}$ are numbers of bosonic and fermionic fields,
respectively. Although Garriga et.al also discussed the semiclassical 
Einstein equation, they gave a solution for the $\Lambda=0$ case
only, which is not relevant for the purpose of this paper. 
Since the coefficient ${\cal N}$ depends on $L$ and $w_0$, the
equation (\ref{eqn:constraint}) introduces a constraint between
$L$ and $w_0$, or between $\lambda$ and
$\bar{\lambda}$~\cite{Garriga}. The constraint should be related to
the renormalization condition adopted in ref.~\cite{GPT}. The results
are summarized in Table~\ref{table:tension}, in which
$\lambda_0\equiv\lambda/6\kappa^{-2}l^{-1}$,
$\bar{\lambda}_0\equiv\bar{\lambda}/6\kappa^{-2}l^{-1}$ and 
${\cal A}=(5/2)(A/3)^{1/5}=O(1)$.


The so called warp factor can be defined by 
$\phi\equiv e^{\alpha(0)}/e^{\alpha(L)}$ and is calculated to be 
%
\begin{equation}
 \phi = \left[\frac{(\lambda/6\kappa^{-2}l^{-1})^2-1}
	{(\bar{\lambda}/6\kappa^{-2}l^{-1})^2-1}\right]^{1/5}
	\label{eqn:warp-factor}
\end{equation}
for ${\cal N}\ne 0$. (For ${\cal N}=0$, we have 
$\phi=e^{\mp l^{-1}L}$.) 
Hereafter, without loss of generality, we can assume that the brane at 
$w=0$ is our world. We call this brane our brane and another
brane at $w=L$ a hidden brane. From (\ref{eqn:warp-factor}), if we
accept fine tuning of brane tensions, $\phi$ can be made arbitrarily
large. Actually, if $N_{b,f}$ satisfy $I<\kappa^2l^{-3}(N_b-N_f)<32I$,
where $I={\cal A}^{-5}[\int_0^{\infty}(1-y^2)^{-4/5}dy]^5$, then there
exists a set of positive $\lambda$ and positive $\bar{\lambda}$ which 
satisfies the constraint given in Table~\ref{table:tension} and which
gives an arbitrarily large warp factor. If $\kappa^2l^{-3}(N_b-N_f)<I$
then then there exist a set of negative $\lambda$ and positive
$\bar{\lambda}$ which satisfies the constraint given in
Table~\ref{table:tension} and which gives an arbitrarily large warp
factor.


We can easily extend the above results to inflating branes by simply
replacing the metric $\eta_{\mu\nu}$ of the four-dimensional Minkowski
metric with that of the four-dimensional deSitter spacetime
$q_{\mu\nu}$. The five-dimensional metric and energy momentum tensor
are given by (\ref{eqn:metric-ansatz}) and
(\ref{eqn:bulk-energy-momentum}) with $\eta_{\mu\nu}$ replaced by
$q_{\mu\nu}$. For example, in the flat slicing, 
%
\begin{equation}
 q_{\mu\nu}dx^{\mu}dx^{\nu} = -dt^2 + e^{2Ht}(dx^2+dy^2+dz^2). 
\end{equation}
Hence, by using $R+20l^{-2}=0$, we can show that 
%
\begin{equation}
 (\alpha')^2 = 
	-\frac{2}{3}l^{-2}{\cal N}e^{5\alpha} +
	H^2e^{2\alpha}+l^{-2}. 
	\label{eqn:1st-integral-inflating}
\end{equation}

We shall obtain an expression for the warp factor without
solving (\ref{eqn:1st-integral-inflating}) explicitly. However, before
doing it, let us consider the easiest case of ${\cal N}=0$. In this
case, it is easy to integrate (\ref{eqn:1st-integral-inflating}) once
more to get 
%
\begin{equation}
 e^{-2\alpha} = l^2H^2\sinh^2[l^{-1}(w-w_0)]. 
\end{equation}
This is the inflating brane solution which was independently found by
Nihei and Kaloper~\cite{Inflating-brane}. For this solution, the
junction conditions at fixed points $w=0$ and $w=L$ give the following 
equations for $w_0$ and $L$. 
%
\begin{equation}
 w_0 = l\tanh^{-1}
	\left[\frac{6\kappa^{-2}l^{-1}}{\lambda}\right], \quad
 L - w_0= l\tanh^{-1}
	\left[\frac{6\kappa^{-2}l^{-1}}{\bar{\lambda}}\right].
\end{equation}
We can now calculate the four-dimensional cosmological constants
$\Lambda_4$ and $\bar{\Lambda}_4$ measured by observers on our brane
and the hidden brane, respectively. These are given by 
%
\begin{equation}
 \Lambda_4 \equiv 3H^2e^{2\alpha(0)}, \quad
 \bar{\Lambda}_4 \equiv 3H^2e^{2\alpha(L)}, \label{eqn:def-Lambda4}
\end{equation}
since Hubble parameters measured by observers on these branes are
$He^{\alpha(0)}$ and $He^{\alpha(L)}$, respectively. Thence, we obtain
the relations 
$\Lambda_4 = \kappa^4\lambda^2/12 - 3l^{-2}$ and 
$\bar{\Lambda}_4 = \kappa^4\bar{\lambda}^2/12 - 3l^{-2}$.

Now let us consider more complicated case of ${\cal N}\ne 0$. In
this case, one may numerically integrate
(\ref{eqn:1st-integral-inflating}) to obtain a solution. However, 
without solving it, we can obtain an analogue of the expression 
(\ref{eqn:warp-factor}) for the warp factor. First, since
(\ref{eqn:1st-integral-inflating}) is a first order ordinary
differential equation, its solution should have one integration
constant, say $w_0$. Next, we can determine $w_0$ and the distance $L$
between two branes by using the junction conditions
(\ref{eqn:Israel}). These are written as 
%
\begin{eqnarray}
 \left(\frac{\lambda}{6\kappa^{-2}l^{-1}}\right)^2 & = & 
	- \frac{2}{3}{\cal N}e^{5\alpha(0)} 
	+ \frac{1}{3}\Lambda_4l^2 + 1, \nonumber\\
 \left(\frac{\bar{\lambda}}{6\kappa^{-2}l^{-1}}\right)^2 & = & 
	- \frac{2}{3}{\cal N}e^{5\alpha(L)} 
	+ \frac{1}{3}\bar{\Lambda}_4l^2 + 1,
	\label{eqn:junction-inflating}
\end{eqnarray}
where $\Lambda_4$ and $\bar{\Lambda}_4$ are defined by
(\ref{eqn:def-Lambda4}). 
It is easily seen that for any value of $H$ (or $\Lambda_4$) the
range of allowed values of brane tensions is essentially the same as
that for the flat brane case. On the contrary, an analogue of
(\ref{eqn:constraint}) should give a constraint between $\lambda$ 
and $\bar{\lambda}$ which depends on $\Lambda_4$. 
Finally, we can obtain a relation between the
four-dimensional cosmological constant on our brane and the warp
factor. Actually, since $\bar{\Lambda}_4=\phi^{-2}\Lambda_4$, from 
(\ref{eqn:junction-inflating}) we obtain 
%
\begin{equation}
 \phi = \left[\frac{(\lambda/6\kappa^{-2}l^{-1})^2
	-\Lambda_4l^2/3-1}
	{(\bar{\lambda}/6\kappa^{-2}l^{-1})^2
	-\phi^{-2}\Lambda_4l^2/3-1}
	\right]^{1/5}. 
\end{equation}
By using this expression, we can discuss the cosmological constant
problem in this model. First, we can see that, if we
accept a fine tuning of the tension $\bar{\lambda}$ of the hidden
brane, then it is in principle possible to obtain a small $\Lambda_4$ 
and a large $\phi$ at the same time. However, because of the
$\Lambda_4$-dependent constraint between brane tensions, once
$\bar{\lambda}$ is fixed, $\lambda$ should be determined by
$\Lambda_4$. Therefore, in order to obtain a small $\Lambda_4$ and a
large $\phi$ at the same time, we need fine tuning of tension of both
branes.


Throughout this paper, semiclassical Einstein's equation with a
negative cosmological constant and conformally invariant bulk matter
fields has been examined in the brane world scenario with the
$S^1/Z_2$ compactification. When numbers $N_{b,f}$ of bosonic and
fermionic fields satisfy $I<\kappa^2l^{-3}(N_b-N_f)<32I$, we
obtained an exact semiclassical solution which has two static
branes with positive tension and for which the warp factor can be
arbitrarily large. When $\kappa^2l^{-3}(N_b-N_f)<I$, we 
have also obtained an exact semiclassical, static solution for which
our brane has negative tension, the hidden brane has positive tension
and the warp factor can be arbitrarily large. However, in order to
obtain a large warp factor, fine tuning of brane tensions is required
for both cases.

In order to justify the fine tuning of brane tensions, we need another
new mechanism. In this sense, so far, the hierarchy problem has not
yet been solved by this model. Besides this issue, we have another
issue to be made clear in future. Namely, it is not clear whether the
large warp factor is sufficient or insufficient to solve the hierarchy
problem, since in this model there appears a peak of the warp factor
between our brane and the hidden brane. If the peak is high enough
then it may be expected that each of two branes behaves like an
isolated single brane~\cite{Tanaka}. Therefore, more detailed analysis
will be required to understand roles of the warp factor in this
model.

Extension to general homogeneous, isotropic branes may be an
interesting future work. When we set $T_{MN}=0$, a system in this 
category should reduce to the cosmological brane solution which was 
independently found by several people~\cite{Friedmann}. Since the
geometry is less symmetric than the inflating brane discussed in this
paper, we cannot expect the great simplification. On the contrary,
since acceleration of branes changes in time, we can expect an 
interesting phenomenon caused by a quantum effect called a moving
mirror effect. It is well known that the moving mirror effect makes
energy momentum tensor dependent of higher derivatives of boundary
positions in a non-local way~\cite{BD}. From results in two-dimension, 
it is expected that corrections due to the moving mirror effect 
vanish if and only if the boundary is static (the flat brane case) or
uniformly accelerating (the inflating or deSitter brane case). Hence,
it seems worth while investigating more general cases in detail.

Analysis of the stability against perturbations seems an important
future work. In particular, the so called radion stability is worth
while investigating. In ref.~\cite{GPT}, Garriga et.al analyzed the
radion stability and concluded that the $N_b>N_f$ case is unstable
while the $N_b<N_f$ case is stable, implicitly assuming that the
kinetic term of the radion field is given classically and that all
quantum effects are included in the potential only. However, as stated
above, the moving mirror effect caused by the motion of branes
introduces non-local, higher-derivative corrections to the energy
momentum tensor. Hence, it is expected that the kinetic term will be
corrected significantly by quantum effects and that the dynamics of
the radion will become non-local if the moving mirror effect is taken
into account. It seems that we need more systematic analysis of the
stability.

\vspace{1cm}

The author would like to thank Professor W.~Israel for continuing
encouragement and Dr. T.~Shiromizu for useful comments. Part of this
work was done during Workshop on String Cosmology in UBC. The author 
would be grateful to Professors R. Brandenberger and W. G. Unruh for
their hospitality. The author is supported by CITA National Fellowship
and the NSERC operating research grant.



%
\begin{table}
\caption{Range of tensions, values of $w_0$ and $L$, and a constraint
between tensions (${\cal A}=(5/2)(A/3)^{1/5}=O(1)$)}
        \label{table:tension}
\begin{center}
\begin{tabular}{|c||c|c|c|c|} 
 $N_b,N_f$ & 
 $\lambda_0\equiv\lambda/6\kappa^{-2}l^{-1},
 \bar{\lambda}_0\equiv\bar{\lambda}/6\kappa^{-2}l^{-1}$ 
 & $w_0$ & $L-w_0$ &  Constraint \\ \hline 
 $N_b>N_f$ & 
 $0 < \lambda_0 \leq 1$, $0 < \bar{\lambda}_0 \leq 1$& 
 $\frac{2}{5}l\tanh^{-1}\lambda_0$ &
 $\frac{2}{5}l\tanh^{-1}\bar{\lambda}_0$ &
 $\int^{\lambda_0}_{-\bar{\lambda}_0}
 (1-y^2)^{-4/5}dy = {\cal A}[\kappa^2l^{-3}(N_b-N_f)]^{1/5}$ \\ 
  & or
	$0 \leq -\bar{\lambda}_0 < \lambda_0 \leq 1$ & 
  & &   \\
  & or
 $0 \leq -\lambda_0 < \bar{\lambda}_0 \le 1$ &
  & &   \\
 $N_b<N_f$ & 
 $1\leq\lambda_0<-\bar{\lambda}_0$&
 $\frac{2}{5}l\tanh^{-1}\lambda_0^{-1}$ &
 $\frac{2}{5}l\tanh^{-1}\bar{\lambda}_0^{-1}$ &
 $\int^{-\lambda_0}_{\bar{\lambda}_0}
 (y^2-1)^{-4/5}dy = {\cal A}[\kappa^2l^{-3}(N_f-N_b)]^{1/5}$ \\ 
  & or
 $1\leq\bar{\lambda}_0<-\lambda_0$ &
  & & \\
 $N_b=N_f$ & 
 $\lambda_0=-\bar{\lambda}_0=\pm 1$ &
 Arbitrary & Arbitrary & $\lambda_0=-\bar{\lambda}_0=\pm 1$ \\
\end{tabular}
\end{center}
\end{table}

\end{document}